\documentclass[preprint,superscriptaddress,preprintnumbers,amsmath,amssymb,prd]{revtex4}
\usepackage{graphicx}

\begin{document}

\thispagestyle{empty}

\title{Graphene may help to solve the Casimir conundrum in indium
tin oxide systems}

\author{
G.~L.~Klimchitskaya}
\affiliation{Central Astronomical Observatory at Pulkovo of the
Russian Academy of Sciences, Saint Petersburg,
196140, Russia}
\affiliation{Institute of Physics, Nanotechnology and
Telecommunications, Peter the Great Saint Petersburg
Polytechnic University, Saint Petersburg, 195251, Russia}

\author{
V.~M.~Mostepanenko}
\affiliation{Central Astronomical Observatory at Pulkovo of the
Russian Academy of Sciences, Saint Petersburg,
196140, Russia}
\affiliation{Institute of Physics, Nanotechnology and
Telecommunications, Peter the Great Saint Petersburg
Polytechnic University, Saint Petersburg, 195251, Russia}
\affiliation{Kazan Federal University, Kazan, 420008, Russia}

\begin{abstract}
We reconsider the long-explored problem that the magnitude of the measured
Casimir  force between an Au sphere and an indium tin oxide (ITO) film
decreases significantly with no respective changes in the ITO dielectric
permittivity required by the Lifshitz theory. Two plausible resolutions
of this conundrum are discussed: the phase transition of an ITO film from
metallic to dielectric state and the modification of a film surface under
the action of UV light. To exclude the latter option, we propose an improvement
in the experimental scheme by adding a graphene sheet on top of an ITO film.
The formalism is developed allowing precise calculation of the Casimir force
between an Au sphere and a graphene sheet on top of ITO film deposited on
a quartz substrate. In doing so Au, ITO, and quartz are described by the
frequency-dependent dielectric permittivities and real graphene sheet with
nonzero mass-gap parameter and chemical potential by the polarization tensor
at nonzero temperature. Numerical computations performed both before and after
the phase transition resulting from the UV treatment show that the presence of
graphene leads to only a minor decrease in the drop of the Casimir force which
remains quite measurable. At the same time, in the presence of graphene the guess
that an observed drop originates from the modification of an ITO surface by the
UV light breaks down. Similar results are obtained for the configuration of two
parallel plates consisting of a graphene sheet, an ITO film and a quartz substrate.
The proposed experiments involving additional graphene sheets may help in
resolution of the problems arising in application of the Lifshitz theory to real
materials.
\end{abstract}

\maketitle

\section{INTRODUCTION}

With advances in microelectronics, the fluctuation-induced van der Waals and
Casimir forces attract the particular attention of many researches.
Taking into account that these forces become dominant at separations below
a few hundred nanometers, they can be used for actuation of various
microdevices in place of the electric force \cite{1,2}.
Work in this scientific direction has already come up with many promising results.
Thus, the stability of Casimir-actuated microdevices depending on geometry
and dielectric properties of materials was considered in Refs.~\cite{3,4}.
The role of surface roughness in the actuation of microelectromechanical
systems by the Casimir force was found \cite{5,6}. Sensitivity of micromechanical
Casimir actuation on amorphous-to-metallic phase transitions was also
determined \cite{7}. Much attention is given to recent experimental demonstration
of the Casimir forces on a silicon micromechanical chip \cite{8,9}.
Moreover, there are suggestions to control mechanical switchers using the Casimir
force between a graphene sheet and a silicon membrane \cite{10} and to devise
an optical chopper driven by the Casimir force \cite{11}.

The wide diversity of promising practical implementations of fluctuation-induced
forces in various microdevices makes highly desirable a clear knowledge of their
nature. The unified theoretical description of the van der Waals and Casimir forces
is given by the fundamental Lifshitz theory \cite{12,13}, which expresses both the
free energy and force via the frequency-dependent dielectric permittivity of
boundary materials. It happened, however, in some cases that a comparison between
the experimental data and theoretical predictions resulted in big surprises.

A serious conundrum was revealed in application of the Lifshitz theory to systems
including indium tin oxide (ITO) films. This semiconductor finds wide applications as an optoelectronic material possessing high electrical conductivity and optical
transparency, but it is opaque in the ultraviolet (UV) frequency region.
Thin ITO films can be easily deposited on different substrates \cite{14}.
Pioneer measurements of the Casimir interaction between an Au sphere and an ITO film
have been performed in Refs.~\cite{15,16}. The gradient of the Casimir force was
measured to be roughly 40\%--50\% smaller than that between an Au sphere and an Au
plate. This result is in qualitative agreement with theoretical predictions of
the Lifshitz theory.

A short time later the Casimir force between an Au-coated polystyrene sphere and
quartz substrate coated with an ITO film was measured in Refs.~\cite{17,18}.
The distinctive property of this experiment is that, after the force measurements
were completed, the ITO sample was UV-treated for 12~h, and then the force
measurements were again performed. Quite unexpectedly, the
measured Casimir force for
the UV-treated samples turned out to be for 21\% and 35\% smaller in magnitude
at 60 and 130~nm sphere-plate separations, respectively, than for the untreated
ones. The observed decrease was not associated with respective modifications
of the optical properties of ITO under the influence of UV treatment.
This was confirmed by means of ellipsometry measurements performed for both
untreated and UV-treated ITO samples. As a consequence, an immediate application
of the Lifshitz theory has led to almost coinciding  Casimir
forces for the untreated and UV-treated ITO samples in clear contradiction with the
experimental evidence.

The hypothetical explanation of the revealed contradiction was provided in
Refs.~\cite{17,18}. It was anticipated that the UV treatment causes the phase transition of the ITO film from metallic to dielectric state.
If, in addition, one assumes that, when calculating the Casimir force, the
contribution of free charge carriers to the ITO dielectric permittivity should
be taken into account in the metallic state, but should be disregarded in the
dielectric state, the theoretical results turn out to be in
good agreement with the
measurement data. The latter assumption is somewhat foreign in the formalism
of the Lifshitz theory, but it was already successfully used for interpretation
of several earlier
experiments. Thus, it was found that the measured difference in the
Casimir forces between an Au sphere and a Si plate in the absence and in the
presence of laser pulse, transforming it from the dielectric to metallic state,
agree with theoretical predictions of the Lifshitz theory only if
free charge carriers in the
dielectric Si are disregarded \cite{19,20}. If free charge carriers
in the dielectric Si
are taken into account, the theoretical predictions are excluded by the data.
One further example is given by the experiment on measuring the thermal
Casimir-Polder force between ${}^{87}$Rb atoms and a SiO${}_2$ plate \cite{21}.
Here, again, the Lifshitz theory was found to be in agreement with the measurement
data if the free charge carriers in the dielectric plate are disregarded
\cite{21,22}, but the same theory  was excluded by the data if the free charge
carriers are taken into account in computations \cite{22}.
It was shown also that inclusion of free charge carriers in calculation of the
Casimir entropy for dielectric materials results in violation of the Nernst heat
theorem, which is satisfied if the free charge carriers are disregarded
\cite{23,24,25,26}.

This raises the question of whether an ITO film really undergoes the phase
transition under the influence of UV treatment or a decrease in magnitude of the
Casimir force for the UV-treated film is due to reducing of some foreign insertions
like dust and contaminants by the action of UV light.
The former is supported by the observation that the UV treatment of ITO leads
to a lower mobility of charge carriers \cite{27}, whereas the latter is based on the
fact that the UV light is often used for cleaning of surfaces \cite{28}, as well as
some other agents (for instance, the Ar-ion cleaning of Au surfaces was recently
used in measurements of the Casimir force \cite{29}). For the resolution of the
above conundrum, it is necessary to assess what is the real impact of UV light on
an ITO film.

In this paper, we propose a minor modification of experiments described in
Res.~\cite{17,18} which allows definite conclusions concerning the
role of UV treatment
of an ITO test body. It is suggested to add a graphene sheet on top of an
ITO film. The Casimir force in graphene systems is a subject of considerable
theoretical study using different calculation approaches (see, e.g.,
Refs.~\cite{30,31,32,33,34}). It was calculated on the basis of first principles
of quantum electrodynamics at nonzero temperature using the polarization tensor
of graphene in (2+1)-dimensional space-time \cite{35,36,37,38,39,40,41,42,43}.
The theoretical results were found to be
in good agreement \cite{44} with measurements
of the gradient of the Casimir force between an Au-coated sphere and
a graphene-coated substrate \cite{45}. Our proposal is based on the fact that
graphene sheets are almost transparent for the UV region \cite{46}.
Thus, the presence of a graphene sheet would leave intact an assumption that
ITO undergoes the phase transition under the influence of UV light, but make
improbable a guess that the observed decrease in the magnitude of the Casimir
force for the UV-treated sample is due to the effect of cleaning.

To confirm the feasibility of the proposed experiment, we calculate the Casimir
force between an Au-coated sphere and a graphene sheet on top of an ITO film
coating the quartz substrate both before and after the UV treatment.
Computations are made using the same values of all parameters as in the
experiments of Refs.~\cite{17,18}.
In so doing, the free charge carriers in the untreated (metallic) ITO are taken
into account and in the UV-treated either taken
into account or disregarded. It is shown that in the presence of graphene the
Casimir force for the UV-treated samples would be from 20\% to 30\%
smaller in magnitude when separation varies from 60 to 150~nm if the free charge
carriers are disregarded. This makes experiment with an additional graphene
sheet quite feasible. We have also computed the Casimir pressure between two
ITO films deposited on two parallel quartz plates
both before and after the UV treatment. Similar to the case of sphere-plate
geometry, it was assumed that the UV treatment results in a phase transition
of ITO from the metallic to dielectric state. According to our computational results,
under an assumption of the phase transition, the Casimir pressure is changed by the
UV light from 56\% to 65\% when separation varies from 150 to 300~nm.
If an additional graphene sheet is added on each plate,
this change varies from 47\% to 52\%, i.e.,
remains easily observable.

The paper is organized as follows. In Sec.~II the general formalism for the
UV treated and untreated ITO films is presented in sphere-plate geometry
in the presence of an additional graphene sheet.
Section~III reports the results of numerical computations for the impact
of graphene sheet on the Casimir force  in sphere-plate geometry.
In Sec.~IV similar results are obtained for the Casimir pressure between two
ITO plates coated with graphene.
In Sec.~V the reader will find our conclusions
and discussion.

\section{The Casimir force in UV-treated and untreated indium tin oxide
system in the presence of graphene}

We consider the configuration of a sphere above a plate. In Refs.~\cite{17,18}
the radius of an Au-coated sphere was $R=101.2~\mu$m. The thickness of an Au
coating was sufficiently large to consider sphere as all-gold.
The sphere-plate separation $a$ exceeds 60~nm. For generality from the very
beginning we consider the plate as a layer structure consisting of a graphene
sheet. an ITO film In${}_2$O${}_3$:Sn of thickness $d=74.6~$nm, and a quartz
substrate of 1~mm thickness. With respect to measurements of the Casimir force,
the latter can be considered as a semispace.

The Casimir force $F_C$ as a function of $a$ at temperature $T$ is given by the
Lifshitz theory where the responses of gold, ITO and quartz to electromagnetic
fluctuations are described by the frequency-dependent dielectric permittivities
and the response of graphene --- by the polarization tensor in (2+1)-dimensional
space-time. In the framework of the proximity force approximation, which is
accurate up to a fraction of a percent for the above parameters \cite{47,48,49,50},
the Lifshitz formula for the layered structure has the usual form
\begin{eqnarray}
&&
F_C(a,T)=k_BTR\sum_{l=0}^{\infty}{\vphantom{\sum}}^{\prime}
\int_{0}^{\infty}k_{\bot}\,dk_{\bot}
\nonumber \\
&&\times\sum_{\alpha}\ln\left[1-
R_{\alpha}^{(p)}(i\xi_l,k_{\bot})r_{\alpha}^{(0,3)}(i\xi_l,k_{\bot})
e^{-2q_la}\right],
\label{eq1}
\end{eqnarray}
\noindent
where $k_B$ is the Boltzmann constant, $\xi_l=2\pi k_BTl/\hbar$ with $l=0,\,1,\,2,\,\ldots$ are the Matsubara frequencies,
$k_{\bot}$ is the magnitude of the project of the  wave vector on the
plane of graphene,
the prime on the summation sign multiplies
the term with $l=0$ by 1/2,   and
\begin{equation}
q_l=\left(k_{\bot}^2+\frac{\xi_l^2}{c^2}\right)^{1/2}.
\label{eq2}
\end{equation}

The summation in $\alpha$ in Eq.~(\ref{eq1}) is by two independent polarizations
of the electromagnetic field,
transverse magnetic ($\alpha={\rm TM}$) and transverse electric ($\alpha={\rm TE}$).
In line with this the TM and TE reflection coefficients of the layered plate are
given by \cite{45}
\begin{equation}
R_{\alpha}^{(p)}(i\xi_l,k_{\bot})=
\frac{r_{\alpha}^{(g)}+r_{\alpha}^{(u)}\left(1\mp
2r_{\alpha}^{(g)}\right)}{1-
r_{\alpha}^{(g)}r_{\alpha}^{(u)}},
\label{eq3}
\end{equation}
\noindent
where the signs minus and plus are for $\alpha={\rm TM}$ and TE, respectively.
Here, the reflection coefficients $r_{\alpha}^{(u)}$ are on a two-layered
structure underlying the graphene sheet, i.e., on an ITO film of thickness $d$
and a thick quartz substrate. It is given by
\begin{equation}
r_{\alpha}^{(u)}(i\xi_l,k_{\bot})=
\frac{r_{\alpha}^{(0,1)}+
r_{\alpha}^{(1,2)}e^{-2k_l^{(1)}d}}{1+
r_{\alpha}^{(0,1)}r_{\alpha}^{(1,2)}e^{-2k_l^{(1)}d}},
\label{eq4}
\end{equation}
\noindent
where the dielectric permittivities of ITO and quartz at the imaginary
Matsubara frequencies are notated as
$\varepsilon_l^{(1)}=\varepsilon^{(1)}(i\xi_l)$ and
$\varepsilon_l^{(2)}=\varepsilon^{(2)}(i\xi_l)$, respectively,
\begin{equation}
k_l^{(n)}=\left[k_{\bot}^2+\varepsilon_l^{(n)}
\frac{\xi_l^2}{c^2}\right]^{1/2},
\label{eq5}
\end{equation}
\noindent
and
 \begin{eqnarray}
&&
r_{\rm TM}^{(n,n')}(i\xi_l,k_{\bot})=
\frac{\varepsilon_l^{(n')}k_l^{(n)}-
\varepsilon_l^{(n)}k_l^{(n')}}{\varepsilon_l^{(n')}k_l^{(n)}+
\varepsilon_l^{(n)}k_l^{(n')}},
\nonumber \\
&&
r_{\rm TE}^{(n,n')}(i\xi_l,k_{\bot})=
\frac{k_l^{(n)}-k_l^{(n')}}{k_l^{(n)}+k_l^{(n')}}.
\label{eq6}
\end{eqnarray}
\noindent
Note that for $n=0$ we have the vacuum gap
$\varepsilon_l^{(0)}=1$.
The reflection coefficient $r_{\alpha}^{(0,3)}$ in Eq.~(\ref{eq1}) is also
given by Eq.~(\ref{eq6}) with $n=0$ and $n^{\prime}=3$ where
$\varepsilon_l^{(3)}=\varepsilon^{(3)}(i\xi_l)$
are the values of the dielectric permittivity of Au at the pure imaginary
Matsubara frequencies.

The reflection coefficients $r_{\alpha}^{(g)}$ on a graphene sheet
in Eq.~(\ref{eq3}) still remain to be defined. They are expressed via the
polarization tensor of graphene in the following way \cite{35,42}
\begin{eqnarray}
&&
r_{\rm TM}^{(g)}(i\xi_l,k_{\bot})=
\frac{q_l\Pi_{00,l}}{q_l\Pi_{00,l}+2\hbar k_{\bot}^2},
\nonumber \\
&&
r_{\rm TE}^{(g)}(i\xi_l,k_{\bot})=-
\frac{\Pi_l}{\Pi_l+2\hbar k_{\bot}^2q_l}.
\label{eq7}
\end{eqnarray}
\noindent
The quantities
$\Pi_{00,l}$ and $\Pi_{{\rm tr},l}=\Pi_{\beta,l}^{\,\beta}$ are the 00
component and the trace of the polarization tensor of graphene, calculated at the
pure imaginary Matsubara frequencies,
\begin{equation}
\Pi_l=k_{\bot}^2\Pi_{{\rm tr},l}-q_l^2\Pi_{00,l}
\label{eq8}
\end{equation}
\noindent
and $q_l$ is defined in Eq.~(\ref{eq2}).

It is convenient to present the explicit expressions separately for
$\Pi_{00,0}$, $\Pi_0$ and for $\Pi_{00,l}$, $\Pi_l$ with $l\geq 1$.
Thus, for $l=0$ one obtains \cite{42}
\begin{eqnarray}
&&
\Pi_{00,0}(k_{\bot})=\alpha\hbar c\frac{k_{\bot}}{v_F}\,
\Psi\!\left(\frac{\Delta}{\hbar v_Fk_{\bot}}\right)+
\frac{8\alpha k_BTc}{v_F^2}\ln\left[\left(e^{\frac{\mu}{k_BT}}+
e^{-\frac{\Delta}{2k_BT}}\right)\left(e^{-\frac{\mu}{k_BT}}+
e^{-\frac{\Delta}{2k_BT}}\right)\right]
\nonumber \\
&&
~
-\frac{4\alpha\hbar c k_{\bot}}{v_F}
\int_{D_0}^{\sqrt{1+D_0^2}}du
\left(
\frac{1}{e^{B_lu+\frac{\mu}{k_BT}}+1}+
\frac{1}{e^{B_lu-\frac{\mu}{k_BT}}+1}\right)
\frac{1-u^2}{\sqrt{1-u^2+D_0^2}},
\nonumber \\
&&
\Pi_{0}(k_{\bot})=\alpha\hbar \frac{v_Fk_{\bot}^3}{c}\,
\Psi\!\left(\frac{\Delta}{\hbar v_Fk_{\bot}}\right)
\label{eq10} \\
&&
~
+4\alpha\hbar\frac{ v_Fk_{\bot}^3}{c}
\int_{D_0}^{\sqrt{1+D_0^2}}du
\left(
\frac{1}{e^{B_lu+\frac{\mu}{k_BT}}+1}+
\frac{1}{e^{B_lu-\frac{\mu}{k_BT}}+1}\right)
\frac{-u^2+D_0^2}{\sqrt{1-u^2+D_0^2}},
\nonumber
\end{eqnarray}
\noindent
Here, $\alpha=e^2/(\hbar c)$ is the fine structure constant,
$v_F\approx c/300$ is the Fermi velocity of the quasiparticles in graphene,
$\Delta$ is the mass-gap parameter, which is usually not equal to zero,
especially for graphene deposited on a substrate \cite{51,52,53},
$\mu$ is the chemical potential describing the fraction of extraneous atoms,
which are always present in graphene \cite{54},
and the following notations are introduced:
\begin{eqnarray}
&&
\Psi(x)=2\left[x+(1-x^2)\arctan\frac{1}{x}\right],
\nonumber \\
&&
D_0=\frac{\Delta}{\hbar v_Fk_{\bot}}, \quad
B_0=\frac{\hbar v_Fk_{\bot}}{2k_BT}.
\label{eq11}
\end{eqnarray}
\noindent

For $l\geq 1$ the exact expressions for $\Pi_{00,l}$ and $\Pi_l$
are somewhat more complicated. However, at $T=300~$K at sufficiently large
separations ($a>60~$nm, as in Refs.~\cite{17,18}) the following approximate
expressions lead to the same results as the exact ones up to a small fraction
of a percent:
\begin{eqnarray}
&&
\Pi_{00,l}(k_{\bot})\approx\alpha\hbar \frac{ck_{\bot}^2}{\xi_l}\,
\left[\Psi\!\left(\frac{\Delta}{\hbar\xi_l}\right)+
\tilde{Y}_l(T,\Delta,\mu)\right],
\nonumber \\
&&
\Pi_{l}(k_{\bot})\approx\alpha\hbar \frac{\xi_lk_{\bot}^2}{c}\,
\left[\Psi\!\left(\frac{\Delta}{\hbar\xi_l}\right)+
\tilde{Y}_l(T,\Delta,\mu)\right],
\label{eq12}
\end{eqnarray}
\noindent
where
\begin{eqnarray}
&&
\tilde{Y}_l(T,\Delta,\mu)=2\int_{\Delta/(\hbar\xi_l)}^{\infty} du
\left(
\frac{1}{e^{B_lu+\frac{\mu}{k_BT}}+1}+
\frac{1}{e^{B_lu-\frac{\mu}{k_BT}}+1}\right)
\nonumber \\
&&~~~~~~~~
\times
\frac{ u^2+\left(\frac{\Delta}{\hbar\xi_l}\right)^2}{u^2+1}
\label{eq13}
\end{eqnarray}
\noindent
and
\begin{equation}
B_l=\frac{\hbar c\tilde{q}_l}{2k_BT}, \quad
\tilde{q}_l=\sqrt{\frac{v_F^2}{c^2}k_{\bot}^2+\frac{\xi_l^2}{c^2}}.
\label{eq14}
\end{equation}

If one puts $r_{\alpha}^{(g)}=0$ in Eq.~(\ref{eq3}), this leads to
$R_{\alpha}^{(p)}=r_{\alpha}^{(u)}$, where $r_{\alpha}^{(u)}$ is defined
in Eq.~(\ref{eq5}), and Eq.~(\ref{eq1}) returns us back to the Casimir
force between an Au sphere and a quartz substrate coated with an ITO film,
as in the experiment of Refs.~\cite{17,18}.

\begin{figure}[b]
\vspace*{-4cm}
\centerline{\hspace*{2.5cm}
\includegraphics{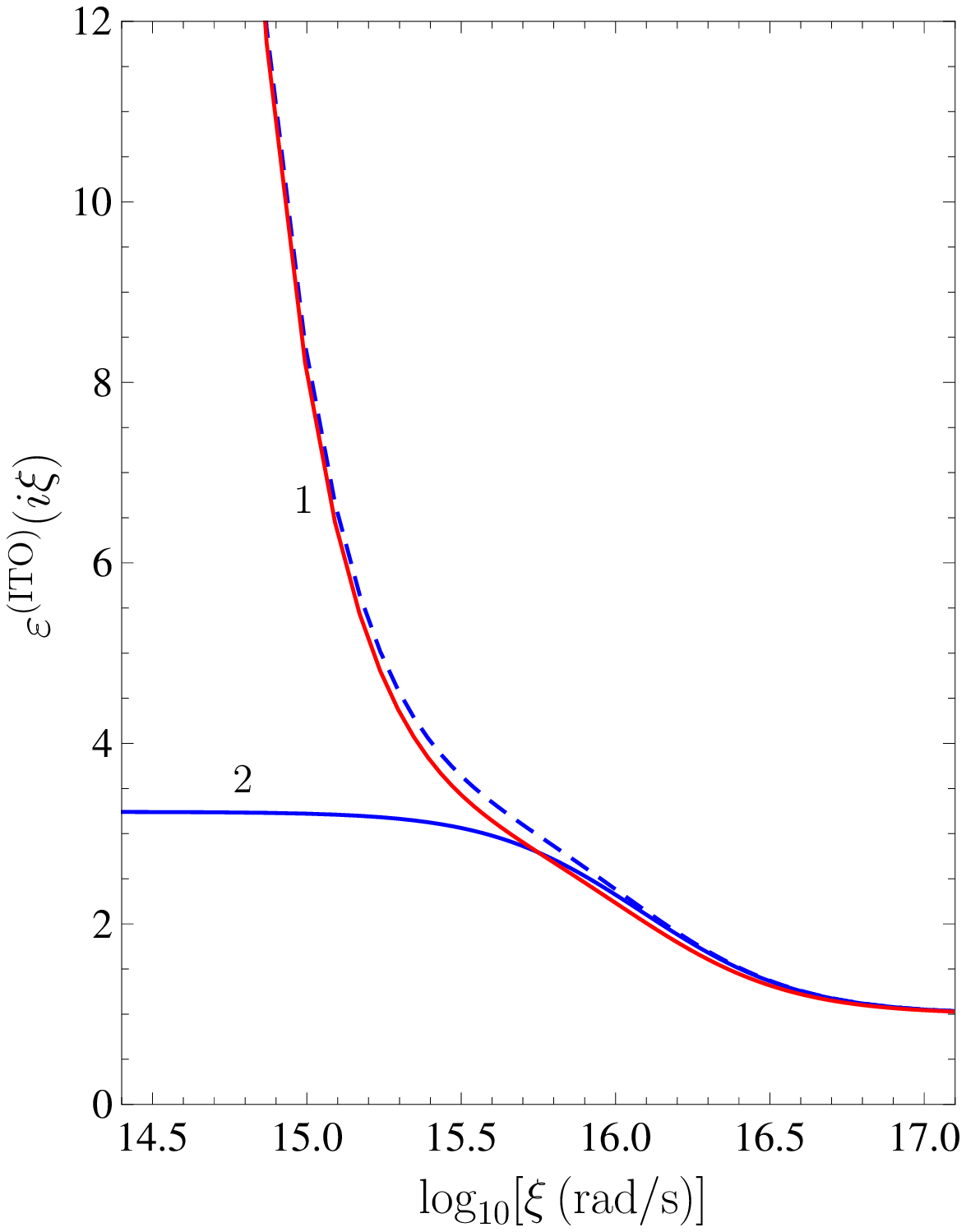}
}
\vspace*{-9.5cm}
\caption{\label{fg1}
The dielectric permittivity of an ITO film as a function of the imaginary
frequency before the UV treatment with account of free charge carriers
(the solid line 1) and after the UV treatment without account (the solid line 2)
and with account (the dashed line) of free charge carriers.
}
\end{figure}

As was mentioned in Sec.~I, measurements of the Casimir force in Refs.~\cite{17,18}
have been performed for both untreated and UV-treated ITO samples.
The UV treatment was performed by means of a mercury lamp with the primary peak
at the wavelength 254~nm. The imaginary parts of the dielectric permittivities were
obtained from ellipsometry measurements performed both before and after UV
treatment. Only minor differences in these imaginary parts were observed
for the untreated and UV-treated samples.
In Fig.~\ref{fg1} we show the dielectric permittivity of ITO as a function of
the imaginary frequency found with the help of Kramers-Kronig relations (see
Refs.~\cite{17,18} for details). The solid lines 1 and 2 demonstrate the
dielectric permittivity before the UV treatment with account of free charge carriers
and after the UV treatment without account of free charge carriers, respectively.
The dashed line shows the dielectric permittivity after the UV treatment, but
with included free charge carriers. The dielectric permittivities of Au and
quartz along the imaginary frequency axis were obtained using the tabulated
optical data \cite{55} and the averaged analytic representation \cite{56} (see
Refs.~\cite{17,18} for details).

\begin{figure}[b]
\vspace*{-9cm}
\centerline{\hspace*{2.5cm}
\includegraphics{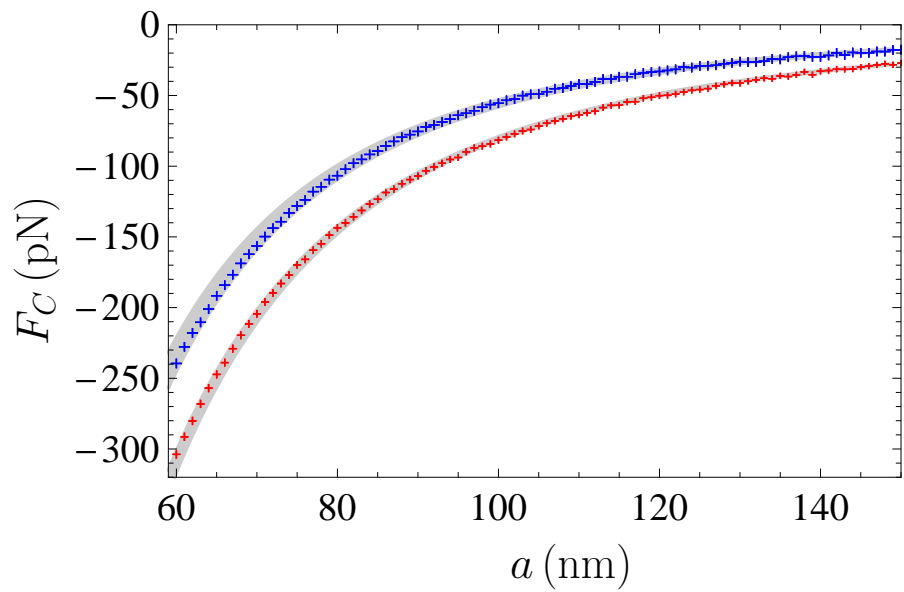}
}
\vspace*{-9.5cm}
\caption{\label{fg2}
The bottom and top bands show the Casimir forces between an Au sphere and
an ITO film deposited on a quartz substrate calculated for the untreated and
UV-treated films, respectively, at $T=300\,$K as functions of separation.
The mean measured Casimir forces are indicated as functions of separation
by the bottom and top sets of crosses for the untreated and
UV-treated films, respectively.
}
\end{figure}

The above formalism (with $r_{\alpha}^{(g)}=0$) leads to two theoretical bands
for the Casimir force $F_C$ between an Au sphere and untreated and UV-treated ITO
film deposited on a quartz substrate shown in Fig.~\ref{fg2}.
The bottom band is computed at $T=300~$K for an untreated film using the solid
line 1 in Fig.~\ref{fg1}, i.e., assuming that ITO is in a metallic state.
The top band is computed at the same temperature for a UV-treated film using the solid
line 2 in Fig.~\ref{fg1} under an assumption that the UV treatment caused the phase
transition from metallic to dielectric state.
Note that the surface roughness measured by means of an atomic force microscope
was taken into account using the geometrical averaging  \cite{18}.
It contributes 2.2\% of the force at $a=60~$nm and less than 1\% at $a\geq 90~$nm.
The widths of the bands are determined
by the theoretical uncertainties mostly connected with the necessity to extrapolate
the ellipsometry data to outside the regions where they are taken.
The mean measured Casimir forces for the untreated and UV-treated samples
are shown together
with their experimental errors determined at a 95\% confidence level by the
bottom and top sets of crosses, respectively \cite{17,18}.
As is seen in Fig.~\ref{fg2}, the theory assuming phase transition as a result of
UV treatment is in a very good agreement with the data. If after the UV treatment
the dielectric permittivity given by the dashed line in Fig.~\ref{fg1} is used in
calculations, the obtained theoretical results almost coincide with the bottom line
in Fig.~\ref{fg2}, i.e., would be in serious contradiction with the measurement data
given by the top set of crosses.

In Fig.~\ref{fg2} it is seen that separation between the two bands exceeds
significantly both the experimental and theoretical errors. In the next section
we show that adding a graphene sheet on top of ITO, which prevents the surface
of ITO from being modified by the UV light, leaves
the theoretical predictions before and
after UV treatment considerably different.

\section{Impact of graphene on the Casimir force in sphere-plate geometry}

Now we assume that there is a graphene sheet on top of ITO film. In this case
the Casimir force between an Au sphere and a layered plate consisting of
graphene, ITO and quartz substrate is given by Eqs.~(\ref{eq1})--(\ref{eq14}).
At first, we find what is the qualitative impact of graphene sheet on the
theoretical bands shown in Fig.~\ref{fg2}. This will be done for a pristine
graphene with $\Delta=\mu=0$. Then we will determine the impact of the mass-gap
parameter $\Delta$ and chemical potential $\mu$.

\begin{figure}[b]
\vspace*{-1cm}
\centerline{\hspace*{2.5cm}
\includegraphics{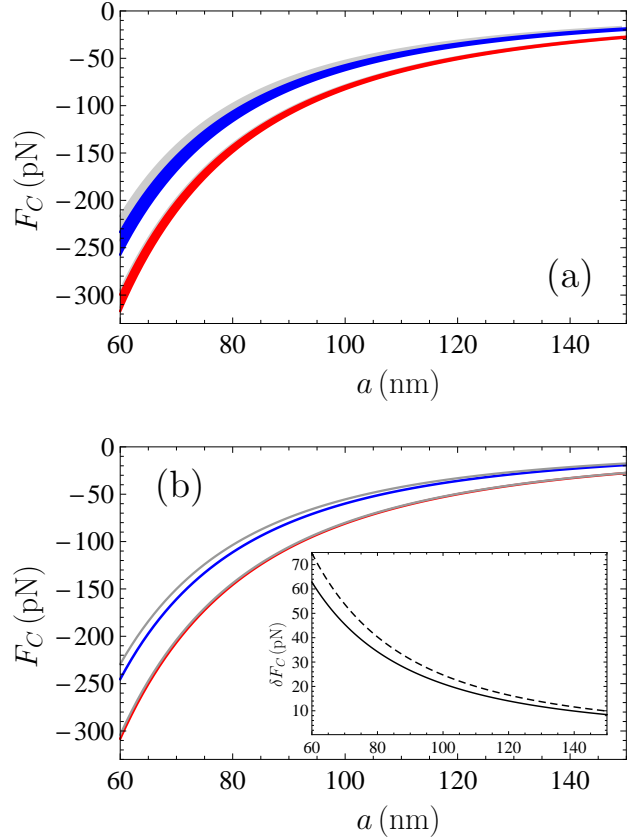}
}
\vspace*{-14.5cm}
\caption{\label{fg3}
(a) The bottom and top dark bands show the Casimir forces between an Au sphere and
a graphene sheet on top of
ITO film deposited on a quartz substrate calculated for the untreated and
UV-treated films, respectively, at $T=300\,$K as functions of separation.
The gray bands demonstrate similar results in the absence of a graphene sheet.
(b) The mean lines of the above bands (see text for further discussion).
In an inset the differences between the dark lines (the UV-treated and untreated
ITO films coated with graphene) and between the gray lines (the uncoated UV-treated and
untreated ITO films) are shown by the solid and dashed curves, respectively.
}
\end{figure}

The computational results for the Casimir force $F_C$ in the presence of a graphene
sheet with $\Delta=\mu=0$ on an ITO film are shown in Fig.~\ref{fg3}(a) by the bottom
and top dark bands as functions of separation for the untreated and UV-treated
samples, respectively. In the same figure the gray bands reproduce that ones
computed in the original experimental configuration  of Refs.~\cite{17,18}, i.e.,
in the absence of a graphene sheet. As in Fig.~\ref{fg2}, the bottom band is
computed using the dielectric permittivity of ITO with account of free charge
carriers (the solid line 1 in Fig.~\ref{fg1}) and the top band is computed with
free charge carriers disregarded  (the solid line 2 in Fig.~\ref{fg1}).
As can be seen in Fig.~\ref{fg3}(a), the presence of a graphene sheet increases
the force magnitudes for a UV-treated sample, but leaves them almost intact for
an untreated one. Computations show that for an untreated sample the presence
of a graphene sheet increases the force magnitude by approximately 1\%.
This cannot be noticed in the figure. Note also that the width of the bands
is determined by the uncertainties in the dielectric permittivity of ITO.
The same widths are obtained for a graphene sheet with nonzero mass-gap
parameter $\Delta\leq 0.1~$eV and chemical potential $\mu\leq 0.5~$eV.

As an illustration, Fig.~\ref{fg3}(b) shows the mean lines of the bands.
The bottom (dark) line almost coincides with the gray one. This shows that
for an untreated (metallic) sample graphene makes only a minor impact on the
Casimir force. At the same time, the medium (dark) line is well away from the
top (gray) line, i.e., an impact of graphene on the UV-treated (dielectric)
sample is quite sensible. As an example, in the presence of graphene the drop
in the magnitude of the Casimir force due to UV treatment varies from 20\% to 35\%
when separation varies from 60 to 150~nm.
An inset in  Fig.~\ref{fg3}(b) shows the differences $\delta F_C$ between the
Casimir forces $F_C^{(g)}$ for UV-treated and untreated ITO films coated
with graphene (the solid line) and
the Casimir forces $F_C$ for uncoated UV-treated and untreated ITO films (the dashed
line). As is seen in the inset, the presence of graphene coating leads to
only a minor decrease in the effect of UV treatment.

The main problem for the feasibility of the proposed experiment is how large is
the force difference between the lower border of the top dark band (for a
UV-treated sample) and the upper border of the bottom dark band (for an
untreated sample) in Fig.~\ref{fg3}(a), as compared to the experimental errors.
For the resolution of this problem
in the case of real graphene sheets with nonzero $\Delta$
and $\mu$, we investigate the quantity
\begin{equation}
F_C^{\rm diff}(a,\mu)=\min\left[F_C^{\rm UV}(a,\mu,\Delta)-
F_C(a,\mu,\Delta)\right].
\label{eq15}
\end{equation}
\noindent
Here, $F_C$ and $F_C^{\rm UV}$ are the Casimir forces for the untreated and
UV-treated samples and the minimum value is taken over the uncertainties in the
dielectric permittivity of ITO and unknown value of the mass-gap parameter of
a graphene sheet in the limits from 0 to 0.1~eV.

\begin{figure}[b]
\vspace*{-9cm}
\centerline{\hspace*{2.5cm}
\includegraphics{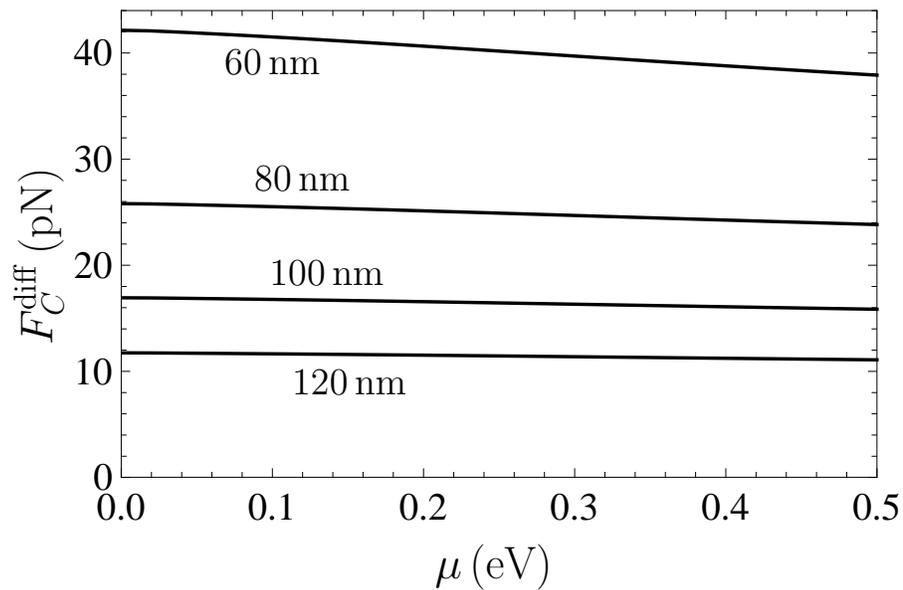}
}
\vspace*{-9.5cm}
\caption{\label{fg4}
The four lines from top to bottom show the minimum differences in the
Casimir forces between an Au sphere and the UV-treated  and untreated
graphene-coated ITO samples
calculated at $T=300\,$K
at separations $a=60$, 80, 100, and 120~nm, respectively,
as functions of the chemical potential.
}
\end{figure}

In Fig.~\ref{fg4} we present the computational results for the quantity
$F_C^{\rm diff}$ by the four lines from top to bottom as
the functions of chemical
potential of graphene at $T=300~$K at the separations $a=60$, 80, 100, and 120~nm,
respectively. As is seen in Fig.~\ref{fg4}, there is an almost linear dependence
of $F_C^{\rm diff}$ on $\mu$ at $a=60~$nm which becomes weaker and weaker with
increasing separation. Thus, at $a=60$, 80, 100, and 120~nm we have
$F_C^{\rm diff}=42.1$, 25.8, 16.9, and 11.7~pN for $\mu=0$, respectively,
and $F_C^{\rm diff}=37.9$, 23.8, 15.8, and 11.1~pN for $\mu=0.5~$eV
at the same respective separation distances.

The obtained results should be compared with the experimental error in measurement
of the quantity $F_C^{\rm diff}$ which is equal to the sum of errors in
measurements of $F_C$ and $F_C^{\rm UV}$. According to the results of
Refs.~\cite{17,18}, at separations of 60, 80, 100, and 120~nm the error in
$F_C^{\rm diff}$ determined at a 95\% confidence level is equal to
5.5, 4.5, 4.0, and 3.5~pN, respectively.
Thus, in spite of the fact that the presence of a graphene sheet somewhat
decreases the differences in theoretical predictions for the Casimir force between
an Au sphere and the untreated and UV-treated ITO films, the feasibility of the
proposed experiment is confirmed.

\section{Impact of graphene on the Casimir pressure between two indium
tin oxide plates}

Taking into consideration
that with recent development of chip technologies \cite{8,9}
the precise Casimir measurements in plane-parallel geometries become
quite realizable, we consider here one more modification of the proposed
experiment gaining an advantage from using two ITO films. For this purpose
we consider the configuration of two parallel plates each of which consists of
a graphene sheet, an ITO film of thickness $d$, and a quartz substrate.
In so doing the graphene sheets facing each other are separated by the distance $a$.
The Casimir pressure in this configuration is given by the Lifshitz formula
\cite{12,13}
\begin{eqnarray}
&&
P_C(a,T)=-\frac{k_BT}{\pi}\sum_{l=0}^{\infty}{\vphantom{\sum}}^{\prime}
\int_{0}^{\infty}q_lk_{\bot}\,dk_{\bot}
\nonumber \\
&&\times\sum_{\alpha}\left[
\frac{e^{2q_la}}{{R_{\alpha}^{(p)}}^2(i\xi_l,k_{\bot})}-1
\right]^{-1}\!\!,
\label{eq16}
\end{eqnarray}
\noindent
where all necessary notations are introduced in Sec.~II.

\begin{figure}[b]
\vspace*{-1cm}
\centerline{\hspace*{2.5cm}
\includegraphics{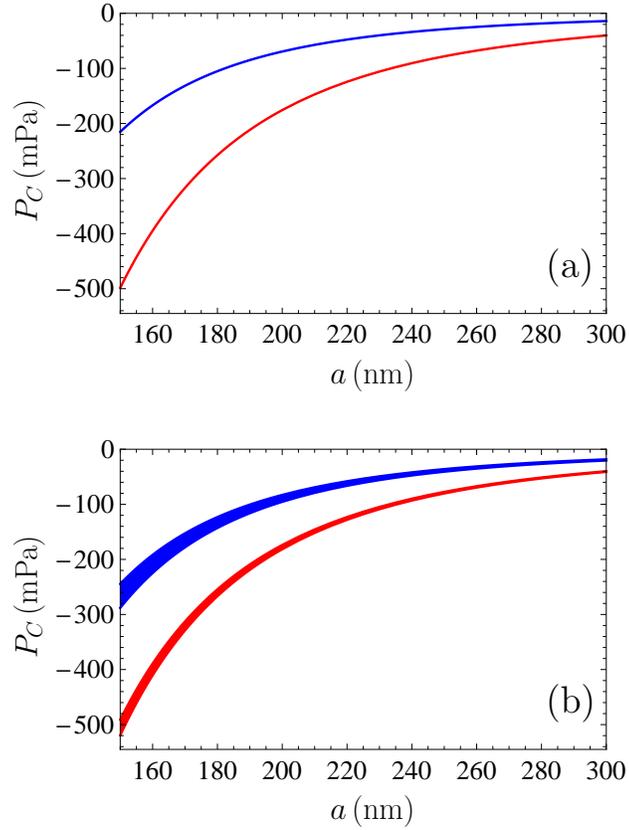}
}
\vspace*{-14.5cm}
\caption{\label{fg5}
(a) The mean values of the Casimir pressure between two plates consisting of
 ITO films deposited on quartz substrates are shown by the bottom and top lines
 computed for the untreated and
UV-treated films, respectively, at $T=300\,$K as functions of separation.
(b) The bottom and top bands show the Casimir pressure in the same conditions,
but for the ITO films coated with additional graphene layers.
}
\end{figure}

First we consider what is the effect of UV treatment for two parallel ITO films
deposited on quartz substrate in the absence of graphene sheets
(i.e., for $r_{\alpha}^{(g)}=0$ in all equations).
In Fig.~\ref{fg5}(a) the mean values of the Casimir pressure computed at $T=300~$K
are plotted as functions of separation by the bottom and top lines for the
untreated and UV-treated ITO samples, respectively.
In the same way as in Sec.~III, the bottom line is computed with the dielectric
permittivity of ITO shown by the solid line 1 in Fig.~\ref{fg1},
and the top line with the dielectric
permittivity shown by the solid line 2 in the same figure.
If after the UV treatment the free charge carriers are taken into account
(i.e.,  the dielectric permittivity shown by the dashed line in Fig.~\ref{fg1}
is used) the obtained Casimir pressure is again presented  to a high accuracy
by the bottom line in Fig.~\ref{fg5}(a).
As is seen in Fig.~\ref{fg5}(a), in the absence of graphene the differences in the
Casimir pressures between the cases of untreated and UV-treated samples vary from
283.3 to 26.2~mPa when separation increases from 150 to 300~nm.
The relative drop in the Casimir pressure due to UV treatment varies from
57\% to 65\%  when separation increases from 150 to 300~nm, respectively.
Here we consider larger separation distances than in Sec.~III because it is common
to determine the Casimir pressure using the dynamic measurement schemes.

Now we assume that there are graphene sheets with $\Delta=\mu=0$ on top of ITO
films and repeat computations of the Casimir pressure (\ref{eq16}) where
$r_{\alpha}^{(g)}$ is given by Eq.~(\ref{eq7}). In  Fig.~\ref{fg5}(b) computational
results are shown as the bottom and top bands where the band widths are caused
by the uncertainties in the dielectric permittivity of ITO.
The bottom band is for an untreated sample and the top band for a UV-treated one,
respectively.
As is seen in Fig.~\ref{fg5}(b), the presence of graphene sheets decreases the
differences in the Casimir pressures in the cases of untreated and UV-treated
samples, but to only a small extent.
Thus, at separations $a=150$, 200, 250, and 300~nm the differences between the
lower border of the top band and the upper border of the bottom band are equal to
203.0, 78.4, 36.8, and 19.6~mPa, respectively. Note that the  nonzero
chemical potential of graphene makes only a minor impact on the obtained results.
In the presence of graphene, the relative drop in the magnitude of the Casimir
pressure due to UV treatment increases from 41\% to 49\%
when separation increases from 150 to 300~nm, respectively.

Since in Sec.~III the same thickness $d$ of an ITO film, as in the experiment
of Refs.~\cite{17,18} was used, here we consider the dependence of the obtained
results on $d$. For this purpose we calculate the quantity
\begin{equation}
P_C^{\rm diff}(a,d)=\min\left[P_C^{\rm UV}(a,d,\mu,\Delta)-
P_C(a,d,\mu,\Delta)\right].
\label{eq17}
\end{equation}
\noindent
Here, $P_C$ and $P_C^{\rm UV}$ are the Casimir pressures for untreated and
UV-treated samples. The minimum value in Eq.~(\ref{eq17})
is taken over the uncertainties in the dielectric permittivity of ITO
and over possible values  of the mass-gap parameter of
graphene $\Delta$ and its chemical potential $\mu$.
Taking into account that with increasing $\mu$ the magnitude of the Casimir pressure
increases, whereas an increase of $\Delta$ leads to the opposite result \cite{42},
we put $\mu=0.5~$eV, $\Delta=0$ in the computations of $P_C^{\rm UV}$ and
$\mu=0$, $\Delta=0.1~$eV in the computations of $P_C$ in order to reach the minimum
value.

\begin{figure}[b]
\vspace*{-9cm}
\centerline{\hspace*{2.5cm}
\includegraphics{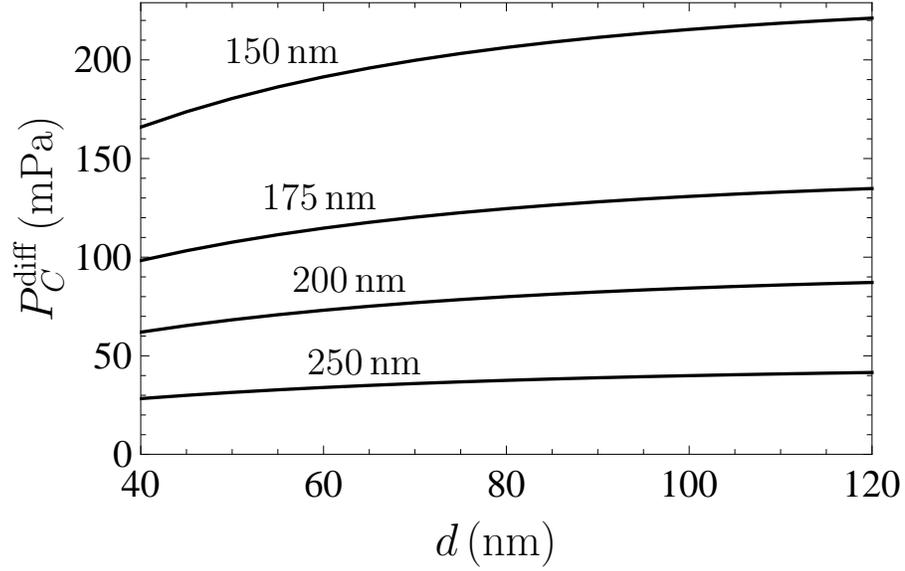}
}
\vspace*{-9.5cm}
\caption{\label{fg6}
The four lines from top to bottom show the minimum differences in
the Casimir pressures between the UV-treated and untreated
graphene-coated ITO samples
calculated at $T=300\,$K
at separations $a=150$, 175, 200, and 250~nm, respectively,
as functions of the thickness of ITO film.
}
\end{figure}

The computational results for $P_C^{\rm diff}$ at $T=300~$K as functions of the
thickness of an ITO film are shown in Fig.~\ref{fg6} by the four lines plotted from
top to bottom at separations $a=150$, 175, 200, and 250~nm between the plates,
respectively.
As is seen  in Fig.~\ref{fg6}, the quantity $P_C^{\rm diff}$ increases rather
slowly with increasing thickness of an ITO film. This increase becomes more rapid
only at the shortest separation considered ($a=150$~nm).
Taking into account that in the dynamic experiments using the atomic force
microscope the effective Casimir pressure is measured with an error of 1~mPa
\cite{57}, an error in the determination of $P_C^{\rm diff}$ could be equal to
2~mPa. It is much smaller than the calculated values of $P_C^{\rm diff}$ shown
in Fig.~\ref{fg6} for any thickness of an ITO film.
This shows that the proposed experiment using two ITO films coated with
graphene sheets is well
suited for observation of the drop in the Casimir pressure under an impact
of UV treatment.

\section{Conclusions and discussion}

In this paper, we have reconsidered the Casimir conundrum in indium tin oxide
systems consisting in the fact that the measured Casimir force drops in magnitude
significantly under the impact of UV treatment with no respective changes in the
dielectric permittivity of ITO, as required by the Lifshitz theory.
In Refs.~\cite{17,18} this drop, which varies from 21\% to 35\% depending on
separation, was hypothetically explained under an assumption that the UV treatment
causes the phase transition of ITO from metallic to dielectric state where
the free charge carriers should not be taken into account similar to two other
experiments of this kind \cite{19,20,21,22}. As an alternative explanation,
one could guess that the UV light modifies an ITO surface by making some kind
of cleaning and this results in respective changes of the Casimir force.

We have proposed a modification in the experimental scheme of  Refs.~\cite{17,18}
by adding a graphene sheet on top of ITO film. This sheet is almost transparent
to the UV light, but prevents the surface of ITO from being modified.
The general formalism for calculation of the Casimir force between an Au sphere
and a graphene sheet on top of ITO film deposited on a quartz  substrate is
developed. In doing so Au, ITO, and quartz are described by the frequency-dependent
dielectric permittivities and graphene by the polarization tensor at nonzero
temperature. Numerical computations using this formalism demonstrate that the
presence of a graphene sheet with different values of the mass-gap parameter and
chemical potential results in only a slight decrease in the drop of the Casimir force
magnitude after the UV treatment of an ITO film. This drop far exceeds the
experimental error confirming the feasibility of the proposed experiment.

As one more proposal, the configuration of two parallel graphene sheets on top
of ITO films deposited on quartz  substrates is considered.
The Casimir pressures are computed both in the absence and in the presence of
graphene sheets before and after the UV treatment. Thus, in the absence of graphene
the drop in the Casimir pressure due to UV treatment varies from 57\% to 65\%
depending on separation. In the presence of graphene sheets this drop, although
becomes somewhat smaller, as yet exceeds the experimental error by the factor
varying from 101 to 10 at different separations between plates. We have also
investigated the dependence of the obtained results on the thickness of an ITO
film and found that it may vary over a wide range without damage to the proposed
experiment.

To conclude, both suggested experiments involving additional graphene layers may
be helpful for the resolution of the Casimir conundrum in indium tin oxide systems
and for further investigation of other unresolved problems in application of the
Lifshitz theory to real materials.

%%%%%%%%%%%%%%%%%%%%%%%%%%%%%%%%%%%%%%%%%%%%%%%%%%%%%%%%%%%%%%%%%%%%%%%%%%%%%%%%
\section*{Acknowledgments}
The work of V.M.M.\ was partially supported by the Russian
Government
Program of Competitive Growth of Kazan Federal University.
%%%%%%%%%%%%%%%%%%%%%%%%%%%%%%%%%%%%%%%%%%%%%%%%%%%%%%%%%%%%%%%%%%%%%%%%%%%%%%%%

%\end{document}
%%%%%%%%%%%%%%%%%%%%%%%%%%%%%%%%%%%%%%%%%%%%%%%%%%%%%%%%%%%%%%%%%%%%%%
\newpage
%%%%%%%__FIGURE__1__%%%%%%%%%%%%%%%%%%%%
%%%%%%%%%%%%%
%%%%%%%__FIGURE__2__%%%%%%%%%%%%%%%%%%%%
%%%%%%%%%%%%%
%%%%%%%%%%%%%%%%%%%%%%%%%%%%%%%%%%%%%%%%%
%%%%%%%__FIGURE__3__%%%%%%%%%%%%%%%%%%%%
%%%%%%%%%%%%%
%%%%%%%__FIGURE__4__%%%%%%%%%%%%%%%%%%%%
%%%%%%%%%%%%%
%%%%%%%__FIGURE__5__%%%%%%%%%%%%%%%%%%%%
%%%%%%%%%%%%%
%%%%%%%__FIGURE__6__%%%%%%%%%%%%%%%%%%%%
%%%%%%%%%%%%%
\end{document}